\documentclass[final,3p,times]{elsarticle}

\usepackage{amssymb}
\usepackage{amsmath}
\usepackage{subcaption}

\journal{Nuclear Physics B}

\begin{document}

\begin{frontmatter}



\title{AMS-L0 silicon ladder assembly with high precision by gantry} 

\author[IHEP]{Feng Wang\corref{cor2}\fnref{label2}}
\author[IHEP]{Baasansuren Batsukh}
\author[IHEP]{Mingyi Dong}
\author[ShanDong]{Cong Liu}
\author[IHEP,UCAS]{Qingze Li}
\author[IHEP]{Sheng Yang}
\author[IHEP]{Congcong Wang}
\author[IHEP,UCAS]{Tianyu Shi}
\author[IHEP,UCAS]{Hengyi Cai}
\author[IHEP,UCAS]{Zixuan Yan}
\author[IHEP]{Jianchun Wang}
\author[IHEP]{Xuhao Yuan\corref{cor1}} 
\cortext[cor1]{xuhao.yuan@ihep.ac.cn}
\cortext[cor2]{wangfeng@ihep.ac.cn}
\affiliation[IHEP]{organization={Institute of High Energy Physics},
            city={Beijing},
            postcode={100049},
            state={Beijing},
            country={China}}
\affiliation[UCAS]{organization={University of Chinese Academy of Sciences},
            city={Beijing},
            postcode={100049},
            state={Beijing},
            country={China}}
\affiliation[ShanDong]{organization={Shandong Institute of Advanced Technology },
            city={Jinan},
            postcode={250300},
            state={Shandong},
            country={China}}
\begin{abstract}
A high-precision silicon microstrip detector assembly methodology based on a gantry system is presented in this paper. The proposed approach has been applied to the mass production of all flight-model detector modules for the L0 tracking detector in the AMS experiment upgrade project, achieving an alignment precision of $3.4~\mu m$ along the strip direction over the maximum length of 1 meter. The fully automated production workflow, integrated with machine vision for accuracy, provides a viable solution for future large-area, high-precision particle detector manufacturing.

\end{abstract}




\end{frontmatter}



\section{Introduction}
\label{sec1}

The Alpha Magnetic Spectrometer (AMS) experiment was jointly developed over 16 years by scientists and engineers from more than 60 research institutions across 16 countries and regions worldwide. AMS is a large-scale cosmic ray detector installed as an independent module on the International Space Station (ISS) in 2011. The first-generation detector, AMS-01, was first tested in orbit aboard NASA's Space Shuttle Discovery in June 1998 \cite{AMS:2002yni}. The upgraded second-generation detector, AMS-02, was successfully deployed in 2011 \cite{Kounine:2012ega}. The detector consists of multiple sub-detectors designed to measure the properties of traversing particles. Its core component is a silicon tracker system composed of nine layers of silicon microstrip sensors. Six of these sensor layers are arranged around a permanent magnet, forming a spectrometer system (inner tracker) capable of determining the charge sign of particles. The Transition Radiation Detector (TRD) located at the top enables discrimination between leptons ($e^{\pm}$) and hadrons (protons and nuclei). A Time-of-Flight (ToF) system measures the direction and velocity of incident particles, as well as their charge. An Anti-Coincidence Counter (ACC) surrounding the tracker within the magnet bore rejects laterally entering particles. A Ring-Imaging Čerenkov (RICH) detector provides high-precision measurements of particle velocity. The Electromagnetic Calorimeter (ECAL), a three-dimensional calorimeter with 17 radiation lengths, is specifically designed for energy measurement of positrons and electrons. Among all current orbital experiments, the AMS remains the only detector capable of distinguishing matter from antimatter, and is expected to maintain this technological uniqueness for the foreseeable future. In January 2020, AMS underwent in-orbit maintenance with the installation of a new cooling system—the Upgraded Tracker Thermal Pump System (UTTPS). This upgrade enables AMS to continue data collection throughout the entire operational lifespan of the ISS, which has recently been extended to 2030.

To further enhance the cosmic ray data statistics during operation and improve the measurement accuracy of nuclear charge and electrical properties, the AMS experiment will implement an upgrade of its top-layer (Layer 0 or L0) upgrade. The new apparatus incorporates two layers of silicon microstrip detector, each structured into 36 independent detector units referred to as “ladder modules”. This enhancement is projected to triple the acceptance across multiple analysis channels while adding two independent dimensions for charge reconstruction. The resulting increase in data statistics will support high-precision spectral analysis of elements with atomic numbers Z = 15 to 30. Through detailed examination of secondary cosmic rays with Z > 14, the upgrade will, for the first time, expose the charge-dependent behavior of cosmic-ray propagation. It will also address the current scarcity of TeV-range spectral statistics in AMS data and establish a foundational empirical basis for a comprehensive theoretical model of cosmic-ray phenomena. These advancements carry significant implications for the field of astrophysics.

In this paper, to ensure alignment accuracy of silicon microstrip ladder over a one-meter span, a novel high-precision assembly process based on a gantry system is developed. This system utilizes machine vision for sensor positioning and a high-accuracy robotic arm for assembly. Furthermore, it incorporates fully automated dispensing functionality, providing a new solution for future high-precision fully automated assembly processes of silicon microstrip detector. Section \ref{sec2} describes the design of the AMS-L0 silicon microstrip detector. Section \ref{sec3} introduces the high-precision assembly system. Section \ref{sec4} elaborates on the high-precision sensor assembly strategy based on the gantry system. Section \ref{sec5} presents the mass production results and discussion of all flight ladders for AMS-L0.

\section{AMS-L0 silicon microstrip detector}
\label{sec2}

\begin{figure}[t]
    \centering
    \begin{subfigure}[b]{0.48\textwidth}
        \includegraphics[width=\textwidth]{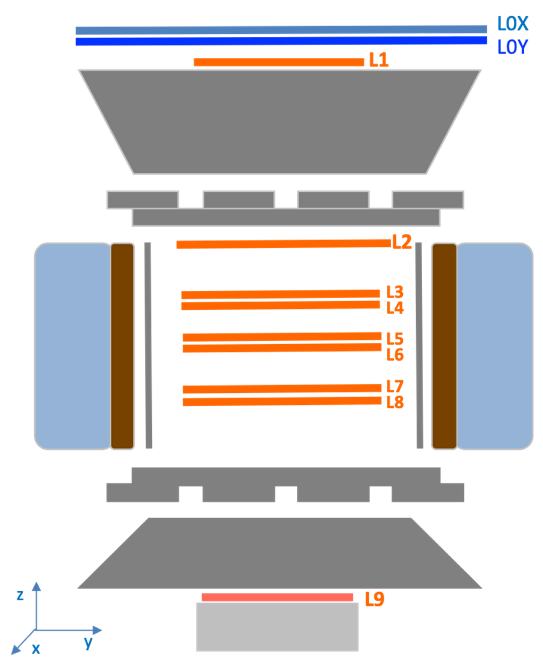}
        \caption{AMS Detector}
        \label{fig:AMSL0-1}
    \end{subfigure}
    \hfill 
    \begin{subfigure}[b]{0.48\textwidth}
        \includegraphics[width=\textwidth]{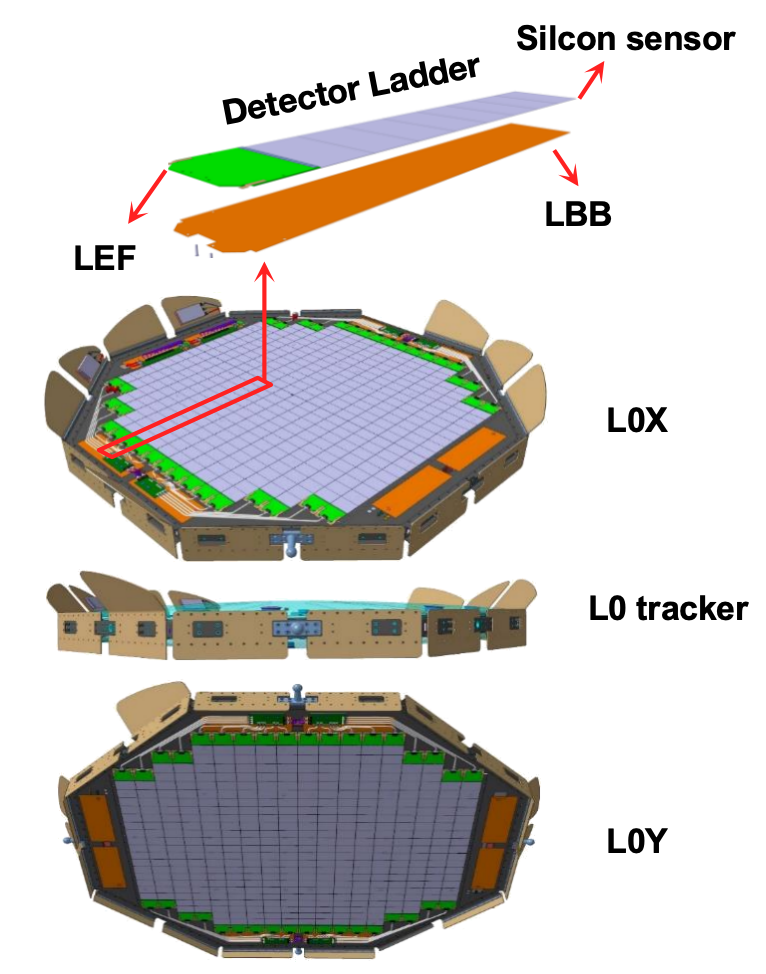}
        \caption{AMS-L0 silicon microstrip detector}
        \label{fig:AMSL0plane}
    \end{subfigure}
    \caption{AMS-L0 upgrade}
    \label{fig:AMSL0}
\end{figure}

\begin{figure}
\centering
\includegraphics[width=0.48\linewidth]{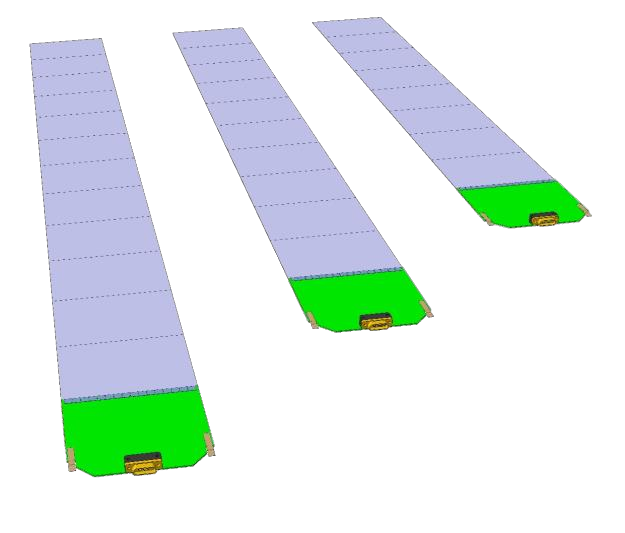}
\caption{\label{fig:ladderlayout}SSD ladder}
\end{figure}

The AMS collaboration plans to upgrade the current detector by installing the L0 detector, which consists of two new silicon microstrip layers, each with an area of approximately 8 $m^2$, on top of the existing apparatus. As shown in Fig. \ref{fig:AMSL0-1}, layers L1 to L9 represent the current silicon tracker planes, while the proposed additional layers, designated L0X and L0Y, will be installed above them. The L0Y layer features strips aligned parallel to the magnetic field direction, enabling measurement of the Y-coordinate. In contrast, the strips in the L0X layer are oriented at a $45^\circ$ angle relative to the magnetic field, allowing simultaneous measurement of both X and Y coordinates, as illustrated in Fig.  \ref{fig:AMSL0plane}. The key of the upgraded detector consists of single-ended readout silicon strip detector (SSD) ladders. Each ladder is constructed by bonding multiple silicon strip sensors end-to-end along the strip direction. Based on the number of sensors per ladder, three types of ladders are defined: 8-SSD, 10-SSD, and 12-SSD, as shown in Fig. \ref{fig:ladderlayout}. 

The core function of the silicon microstrip detector is to precisely measure the trajectory positions of charged particles. Its spatial resolution directly depends on both the strip pitch and the assembly accuracy. The strip pitch of the AMS-L0 sensor is 27.25 $\mu m$ \cite{AMS—L0:sensor}. Key parameters of the silicon sensors are summarized in Table  \ref{tab:SSDladder}. If the assembly accuracy worse than 10 $\mu m$, the uncertainty in charge sharing when a particle crosses the boundary between adjacent strips increases, thereby degrading the position reconstruction accuracy. Furthermore, since AMS-L0 is a detector composed of multiple sensors, cumulative assembly misalignment can significantly affect the spatial consistency of the entire detector. Therefore, to achieve a position resolution of 10 $\mu m$, the alignment accuracy of the detector along the pitch direction must be better than 10 $\mu m$.


\begin{table}[t]
\centering
\begin{tabular}{l|l|l|l|l r}
Parameter &  Rating   &  Unit\\\hline
Device type &  Single-sided AC readout & - \\
Silicon type &  n-type Phosphorus-doped & -\\
Thickness & 320 ± 15 & $\mu m$\\
Chip size &  113,000 ± 20 × 80,000 ± 20 & $\mu m$\\
Strip pitch & 27.25 & $\mu m$\\
Strip width & 10 & $\mu m$\\
\end{tabular}
\caption{\label{tab:SSDladder}Silicon strip sensor parameters}
\end{table}

\section{Gantry system}
\label{sec3}
\begin{figure}[t]
\centering
\includegraphics[width=0.48\linewidth]{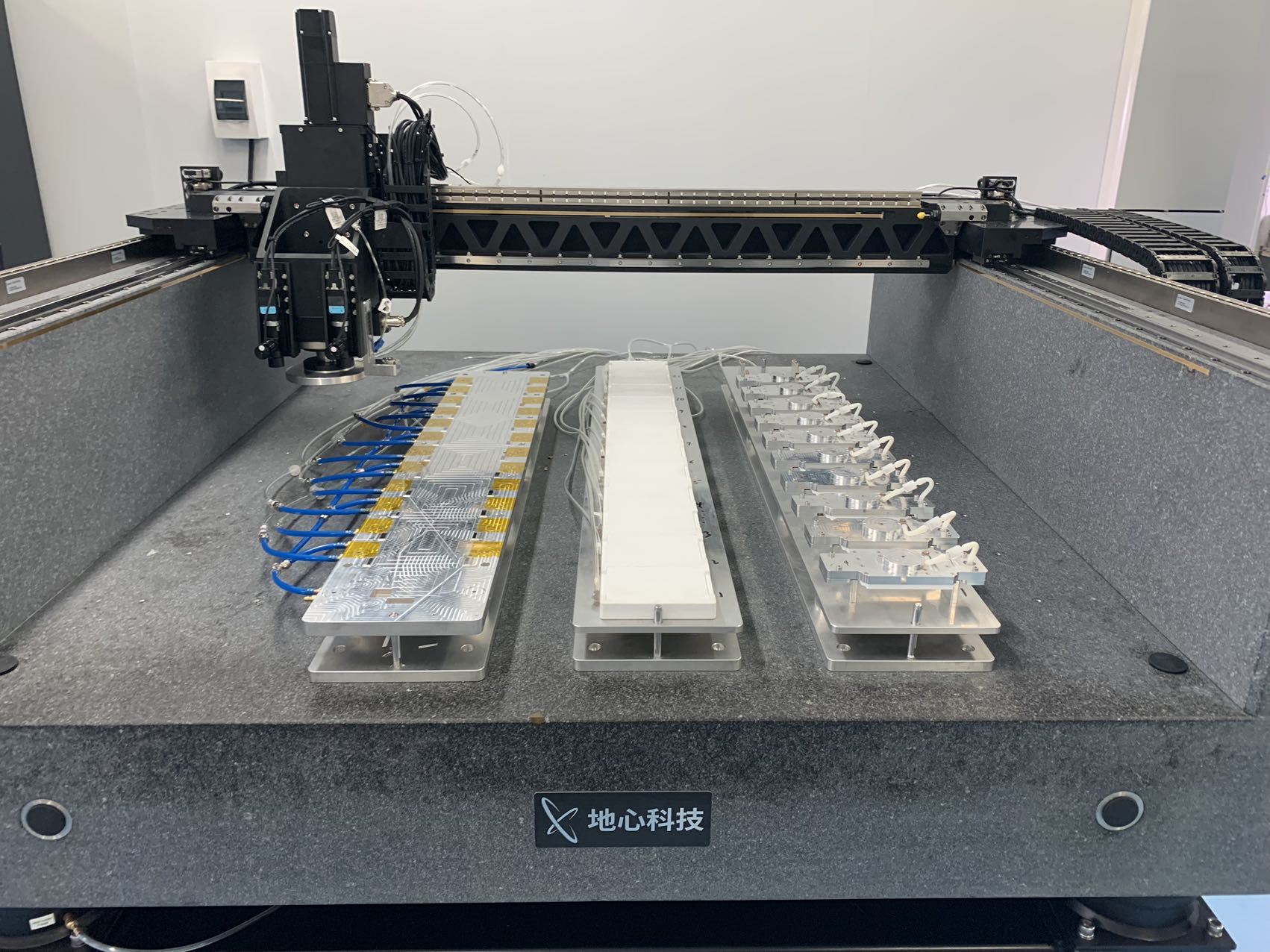}
\caption{\label{fig:gantry_layout}Gantry}
\end{figure}

\begin{figure}[t]
    \centering
    \begin{subfigure}{0.48\textwidth}
        \includegraphics[width=\linewidth]{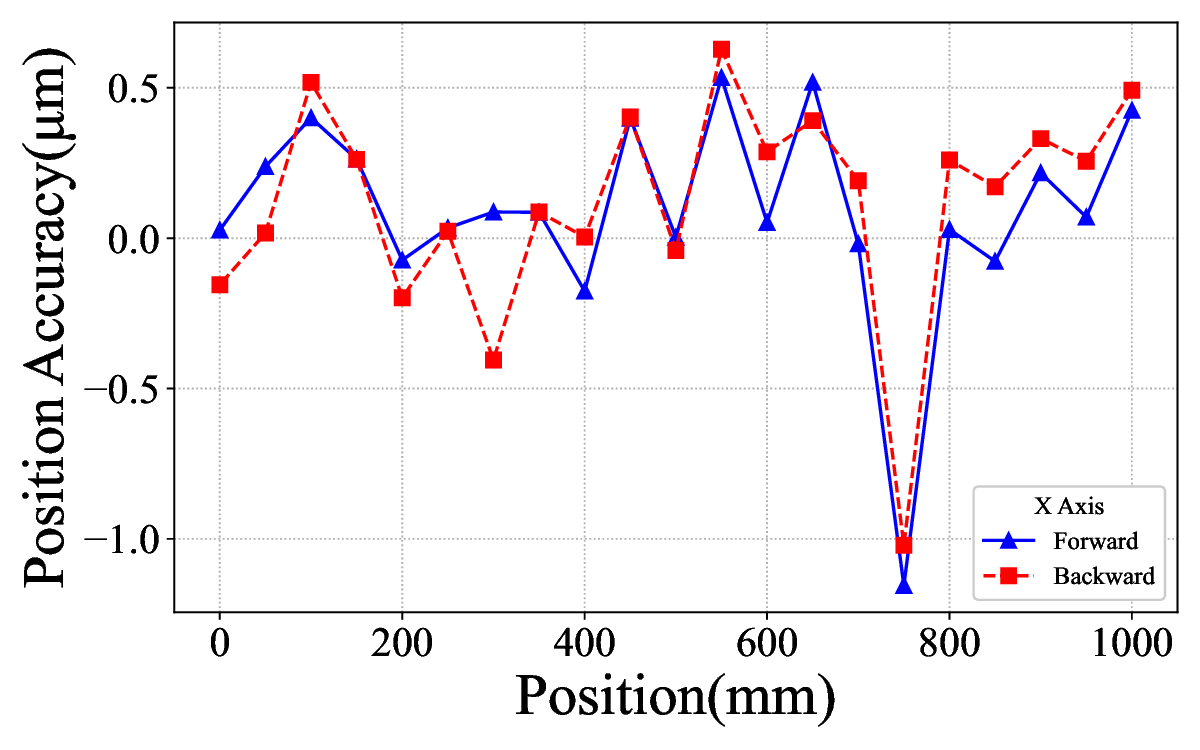}
        \caption{X axis position accuracy}
        \label{fig:X}
    \end{subfigure}
    \hfill
     \begin{subfigure}{0.48\textwidth}
        \includegraphics[width=\linewidth]{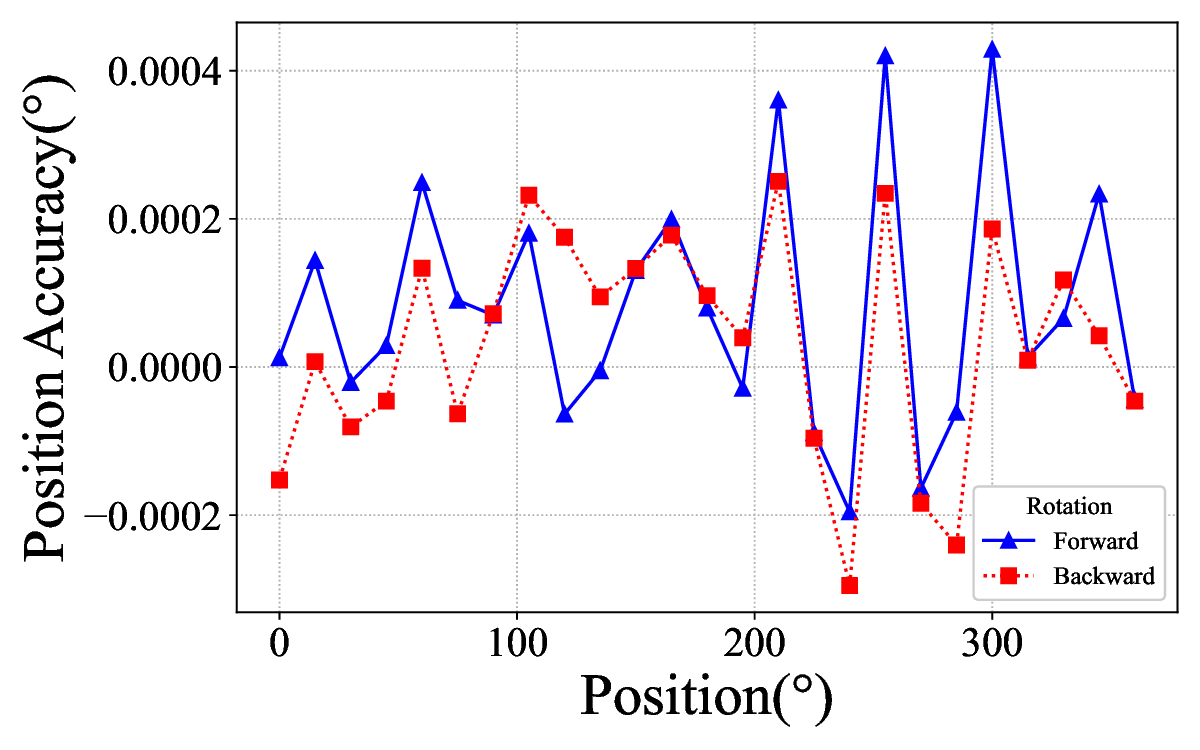}
        \caption{U axis rotation accuracy}
        \label{fig:U}
    \end{subfigure}
    \caption{gantry calibration}
    \label{fig:gantry cali}
\end{figure}

To achieve micron-level assembly precision, we propose utilizing a high-precision gantry system\footnote{Produced by DiXin Technology and customized according to L0 upgrade requirements} integrated with machine vision for the assembly of AMS-L0 ladder. A similar assembly scheme was first implemented in the CMS tracker upgrade project and is currently being applied in the production of modules for the CMS high-granularity calorimeter \cite{Lenzi:2002mda,CMSHGCAL:2020dnm}. The design of the high-precision gantry platform is illustrated in Fig. \ref{fig:gantry_layout}. It comprises four motion axes: X and Y axes for horizontal movement, a Z axis for vertical movement, and a rotational U axis mounted below the Z axis. Equipped with optical encoders for positioning, each axis of the gantry exhibits high positioning accuracy. The positioning accuracy and repeatability of a 1-meter linear axis were measured using a Renishaw laser interferometer. Measurements were taken at 50 $mm$ intervals from 0 to 1000 $mm$, followed by a reverse measurement from 1000 $mm$ back to 0. This procedure yielded the unidirectional positioning accuracy and repeatability over the full 1-meter range, as shown in Fig. \ref{fig:X}. A similar method was applied to evaluate the rotational accuracy and repeatability of the U axis (Fig. \ref{fig:U}). The results demonstrate a linear axis positioning accuracy better than 1 $\mu m$ over 1 meter, and a rotational accuracy of the U axis better than $0.001^\circ$.

The machine vision system is fully integrated with the gantry. It consists of a camera and corresponding image processing algorithms. The camera is mounted on the Z-axis of the gantry, enabling coordinated movement with the XYZ axes. By performing visual recognition of components placed on the gantry worktable, the system provides feedback on the precise position of each part within the gantry coordinate system. With this vision system, the positioning accuracy of components on the gantry can achieve better than 2 $\mu m$. Furthermore, to meet ladder construction requirements, the gantry is also equipped with integrated vacuum and dispensing systems. In this way, fully automated control of the ladder assembly process can be achieved.

\section{AMS-L0 ladder assembly}
\label{sec4}

\begin{figure}[t]
\centering
\includegraphics[width=0.48\linewidth]{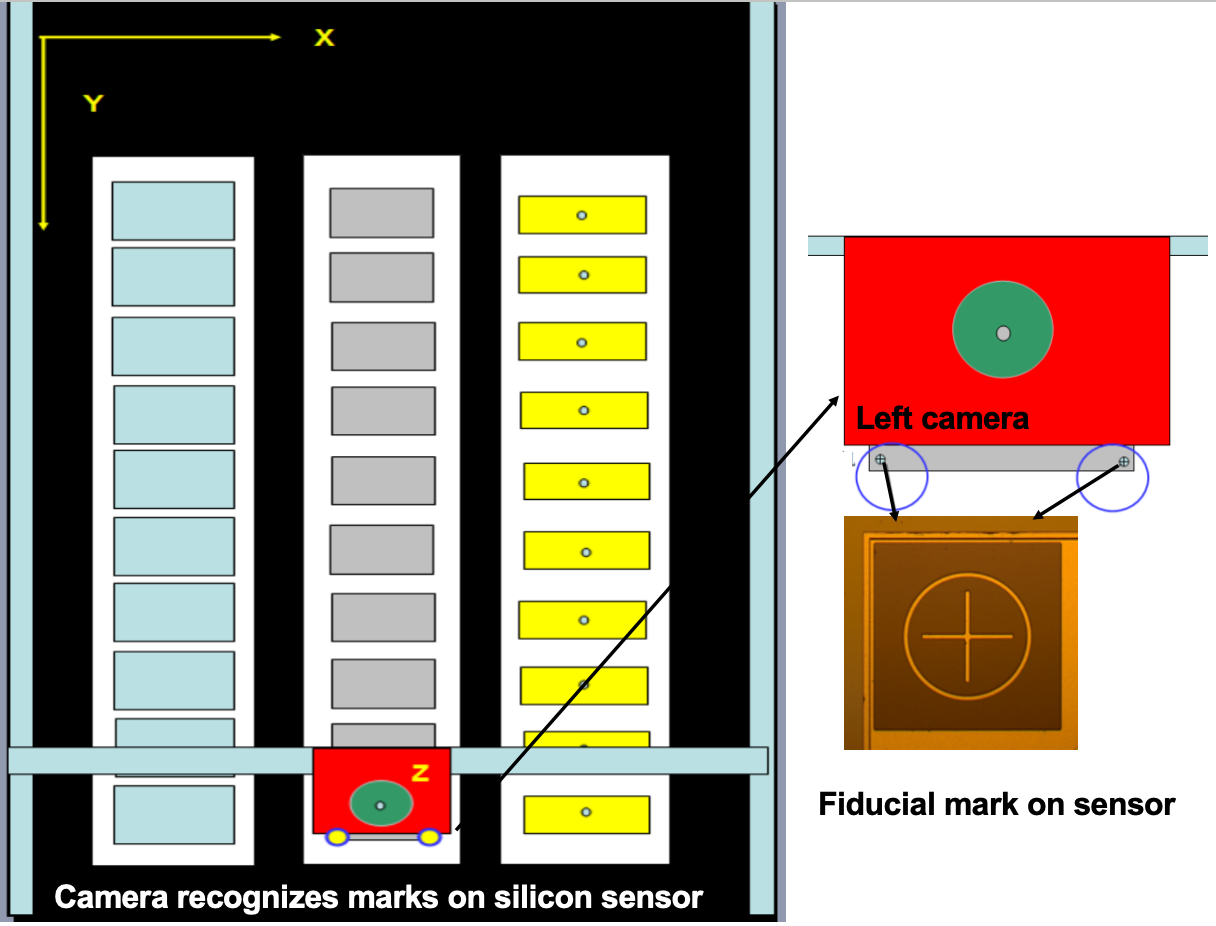}
\caption{\label{fig:gantry_locating}Gantry top-view schematic: camera performing precision measurement of silicon sensor}
\end{figure}

\begin{figure}[htbp]
    \centering
    \begin{subfigure}{0.48\textwidth}
        \includegraphics[width=\linewidth]{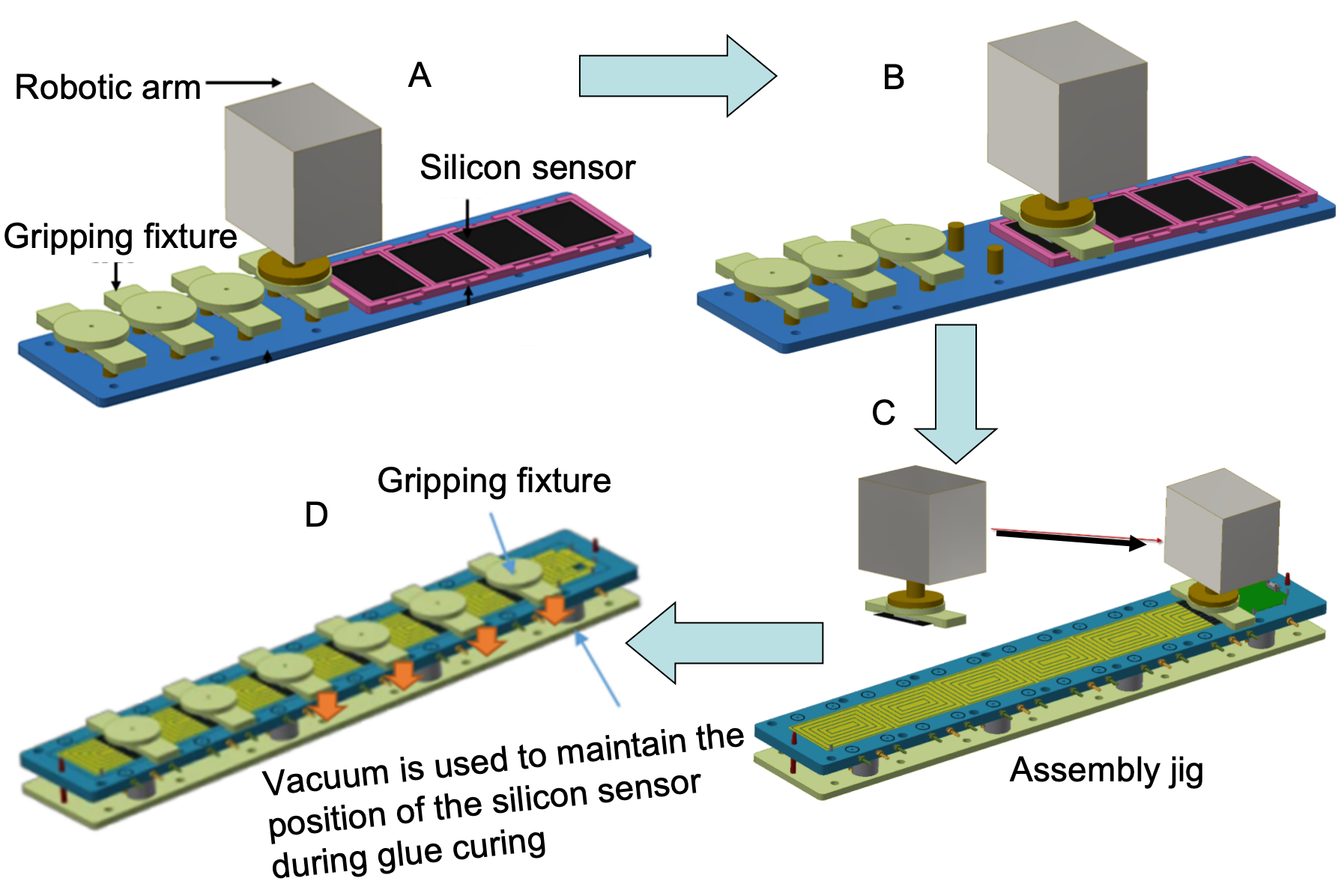}
        \caption{Schematic diagram of the assembly process}
        \label{fig:gantry_assembly}
    \end{subfigure}
    \hfill
     \begin{subfigure}{0.48\textwidth}
        \includegraphics[width=\linewidth]{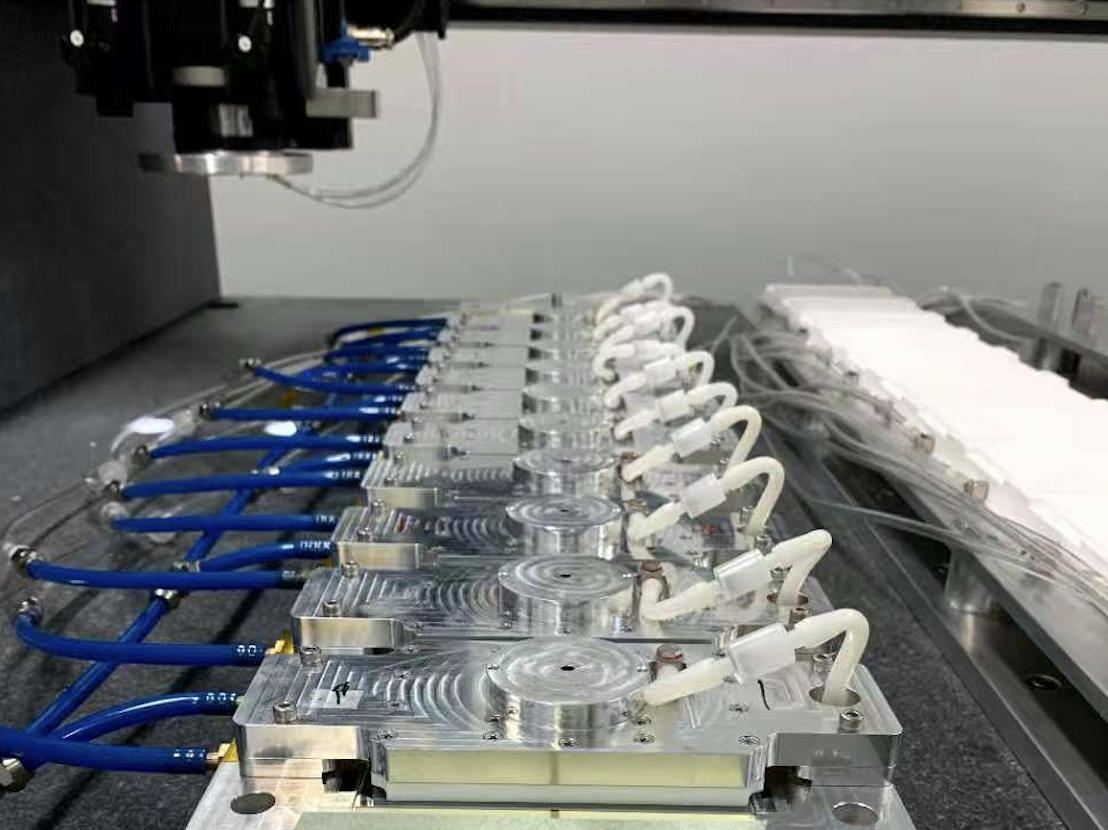}
        \caption{Physical photo of the ladder during curing}
        \label{fig:ladder_assembly}
    \end{subfigure}
    \caption{Gantry Assembly Process Schematic}
    \label{ladderassembly}
\end{figure}

In ladder assembly on the gantry, the positions of the silicon sensors and the assembly areas are identified using machine vision. These positional parameters serve as input to enable the gantry system to automatically move to predefined locations for picking and placing the silicon sensors. Specifically, as illustrated in Fig. \ref{fig:gantry_locating}, the worktable of the gantry is divided into three zones: the left zone is for detector module assembly, the central zone holds silicon sensor, and the right zone contains custom gripping fixtures. During detector assembly, the components and fixtures must be placed in their corresponding zones. Once the program is initiated, the robotic arm of the gantry moves to pre-configured positions to measure the accurately locate both the placement of components and the target positions for ladder assembly. Then, the gantry’s robotic arm sequentially moves above the gripping fixture, engages it by suction (label A in Fig. \ref{fig:gantry_assembly}), then moves to a position directly above the silicon sensor (label B in Fig. \ref{fig:gantry_assembly}), where it picks up the sensor by attaching it to the lower surface of the fixture. Finally, the arm transports both the fixture and the sensor to the position aligned with fiducial markers on the assembly tray. This process is repeated until all sensors are mounted onto the ladder, as shown in Fig. \ref{fig:ladder_assembly}.

\begin{figure}[htbp]
\centering
\includegraphics[width=0.48\linewidth]{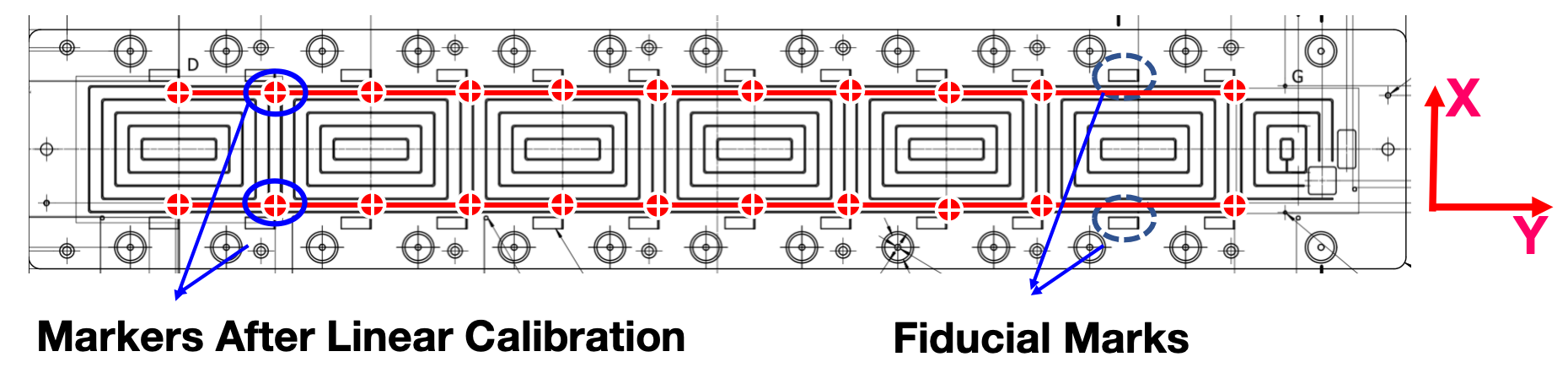}
\caption{\label{fig:assemblytray}Assembly jig drawing}
\end{figure}

\begin{figure}[htbp]
\centering
\includegraphics[width=0.48\linewidth]{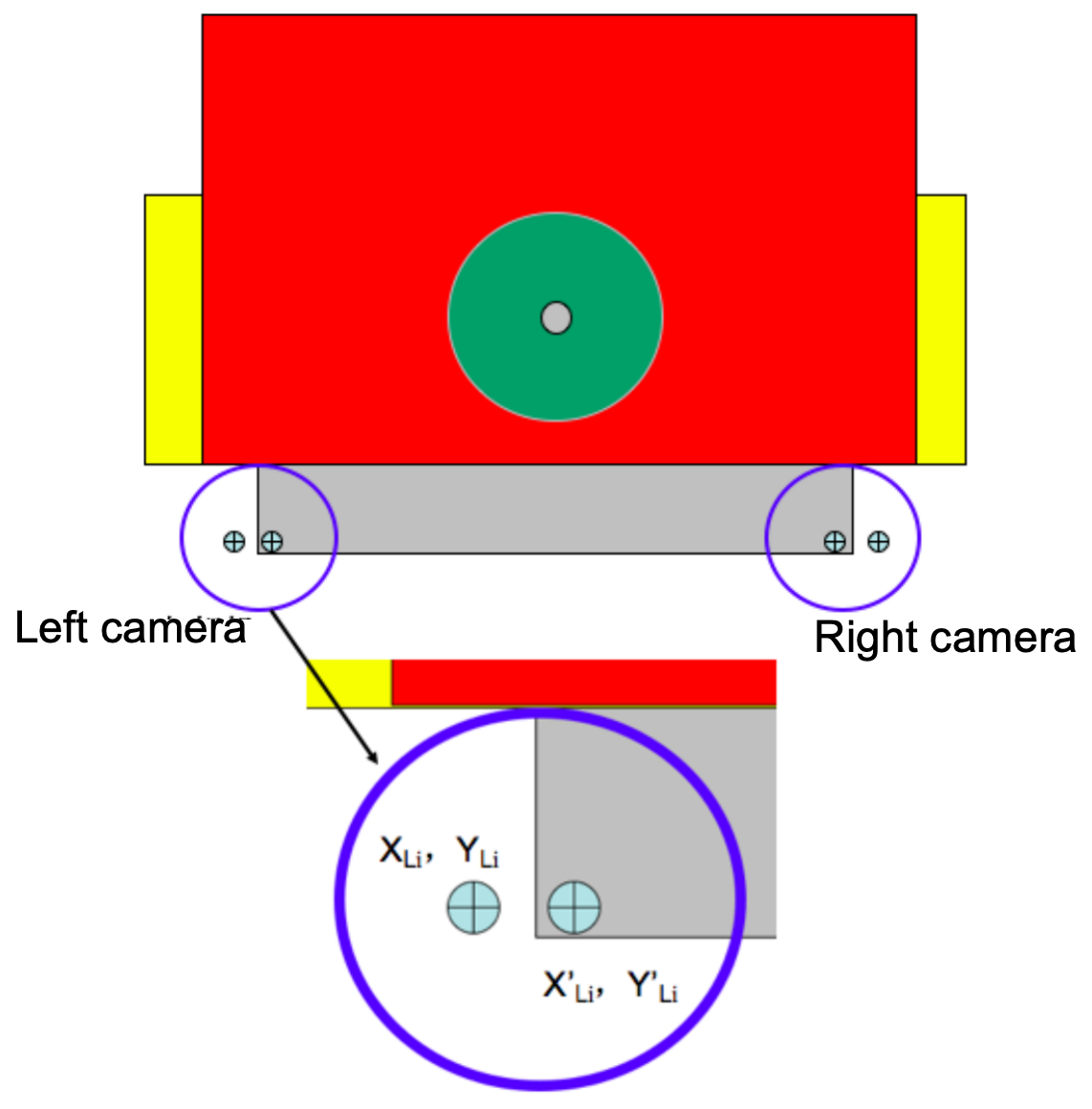}
\caption{\label{fig:putdown}Vision-assisted alignment schematic}
\end{figure}

To address the technical challenge of alignment accuracy during the assembly process, we fabricated approximately symmetric fiducial markers on both sides of the assembly jig used for alignment reference, as shown in Fig. \ref{fig:assemblytray}. The rectangular area between the markers serves as the assembly zone. Due to limitations in machining precision, the positions of these markers may exhibit significant deviations along the strip direction (Y-direction). After fabricating the markers, they are linearly calibrated, resulting in two sets of reference points, $(X_{Li}, Y_{Li})$ and $(X_{Ri}, Y_{Ri})$. In this way, the precision of reference markers is no longer constrained by mechanical processing limitations. When the gantry’s robotic arm picks up the sensor and moves to the $i$-th assembly position, both the left and right cameras simultaneously capture the fiducial markers on the jig and the sensor itself. The real-time coordinates of these markers are acquired as $(X'_{Li}, Y'_{Li})$ and $(X'_{Ri}, Y'_{Ri})$, respectively, as illustrated in Fig. \ref{fig:putdown}. Once the measured sensor positions relative to the reference fiducial markers satisfy the condition given in Equation \ref{eq:1} , the sensor along with the gripping tool is secured onto the assembly jig using vacuum.


\begin{figure}
    \centering
    \includegraphics[width=0.48\linewidth]{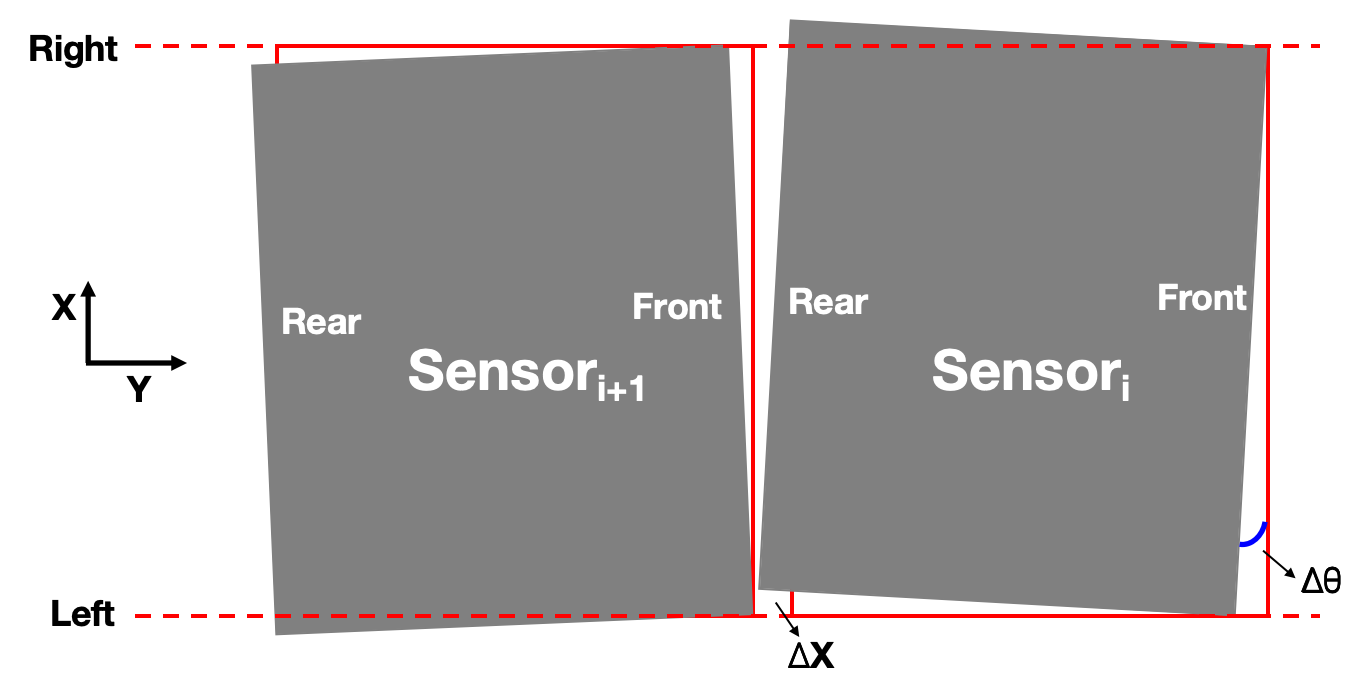}
    \caption{Impact of angular misalignment on X-direction accuracy}
    \label{fig:theta}
\end{figure}

\begin{figure}
    \centering
    \includegraphics[width=0.48\linewidth]{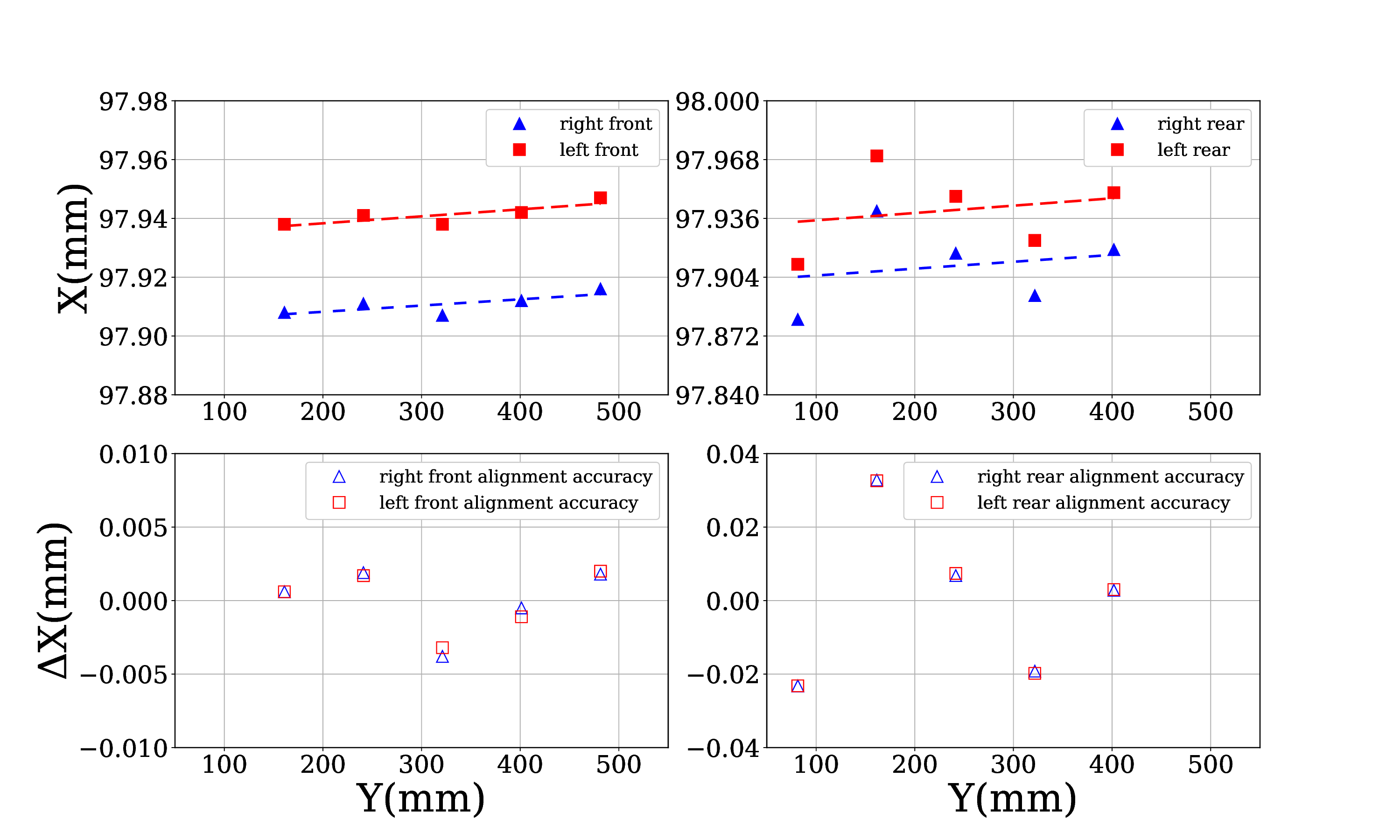}
    \caption{Assembly accuracy for five dummy sensors}
    \label{fig:theta_uncertenty}
\end{figure}

\begin{equation}
|\Delta X_{Li}- \Delta X_{Ri}|<\epsilon, |\Delta Y_{Li} - \Delta Y_{Ri} |<\epsilon,
\label{eq:1}
\end{equation}
where
\begin{align*}
\Delta X_{Li} &= X'_{Li} - X_{Li} \\
\Delta X_{Ri} &= X'_{Ri} - X_{Ri} \\
\Delta Y_{Li} &= Y'_{Li} - Y_{Li} \\
\Delta Y_{Ri} &= Y'_{Ri} - Y_{Ri}
\end{align*}
Equation \ref{eq:1} incorporates alignment constraints not only in the X-direction but also in the Y-direction. Restricting alignment to the X-direction alone could introduce rotational misalignment of the sensor during assembly. Such rotational error would degrade the X-direction alignment precision at the rear side of the sensor, as illustrated in Fig. \ref{fig:theta}. The alignment of the 5-SSD dummy ladder is shown in  Fig. \ref{fig:theta_uncertenty}. It was clearly observed that without ensuring angular alignment accuracy, the alignment precision of the front-left and front-right fiducial markers could be maintained within 5 $\mu m$, while the deviation at the rear-left and rear-right markers could be worse than 40 $\mu m$. The effect of angular misalignment on positioning accuracy can be estimated using Equation \ref{eq:2}, where L represents the width of the sensor along the strip-direction. Therefore, we strictly control the angular accuracy to be better than $0.005^\circ$ during the assembly process.


\begin{equation}
\Delta X= L*tan\theta
\label{eq:2}
\end{equation}

\begin{figure}[t]
    \centering
    \includegraphics[width=0.48\linewidth]{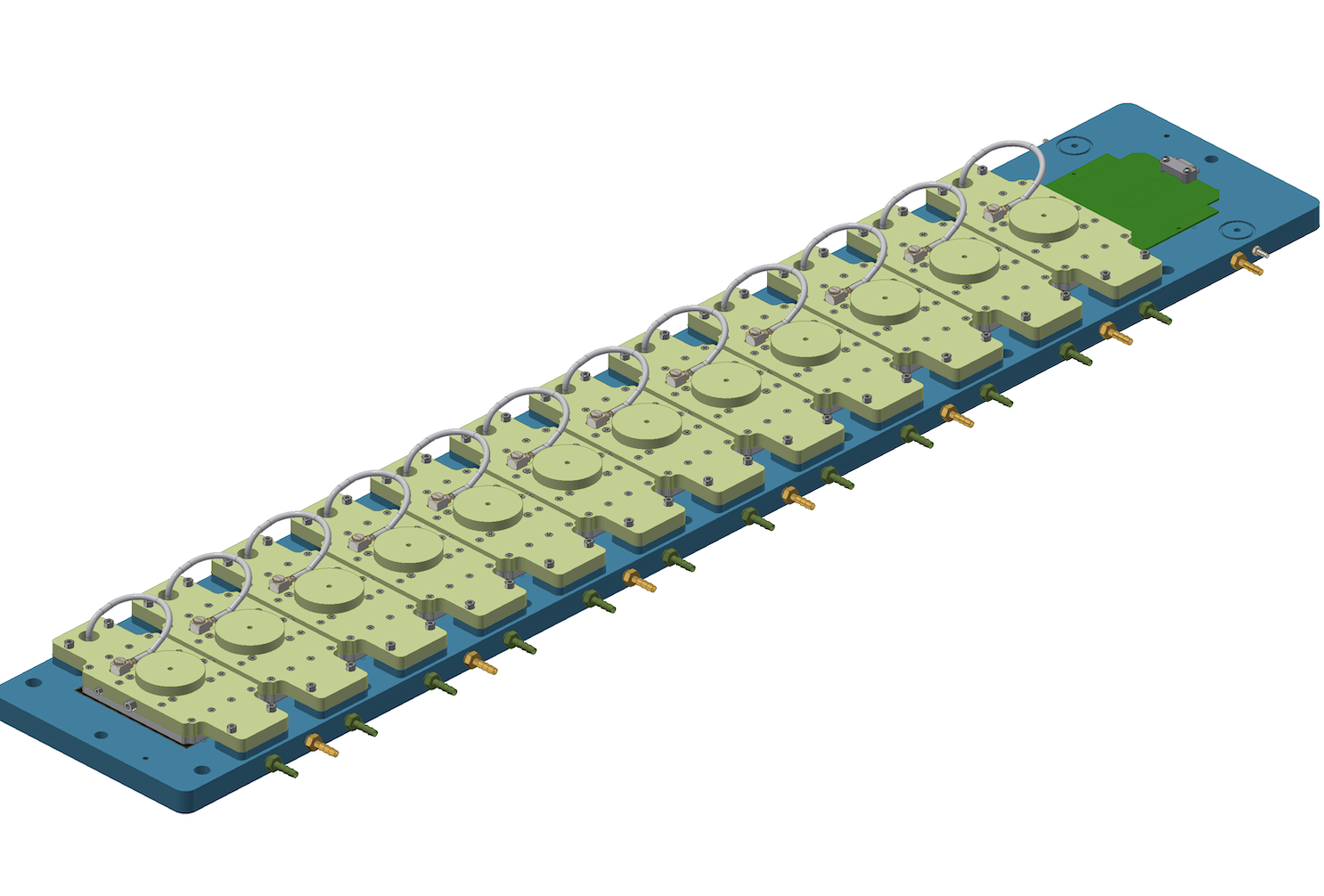}
    \caption{Vacuum suction fixture}
    \label{fig:pick-up-tool}
\end{figure}

To prevent displacement of the sensor during the curing of structural adhesive, we designed a vacuum suction fixture \cite{Zhuanli}, as shown in Fig. \ref{fig:pick-up-tool}. After completing the aligned placement of the sensor, vacuum is applied at the corresponding positions of the assembly jig to secure the suction fixture. Simultaneously, the base of the suction fixture features an integrated vacuum channel at the contact interface with the assembly jig. While securing the fixture, this channel enables the vacuum to firmly hold the sensor against the lower surface of the suction fixture. Thus, during the curing process, the sensor and suction fixture act as an integrated unit, immobilized under vacuum at the aligned position without any movement.

\begin{figure}[t]
    \centering
    \includegraphics[width=0.48\linewidth]{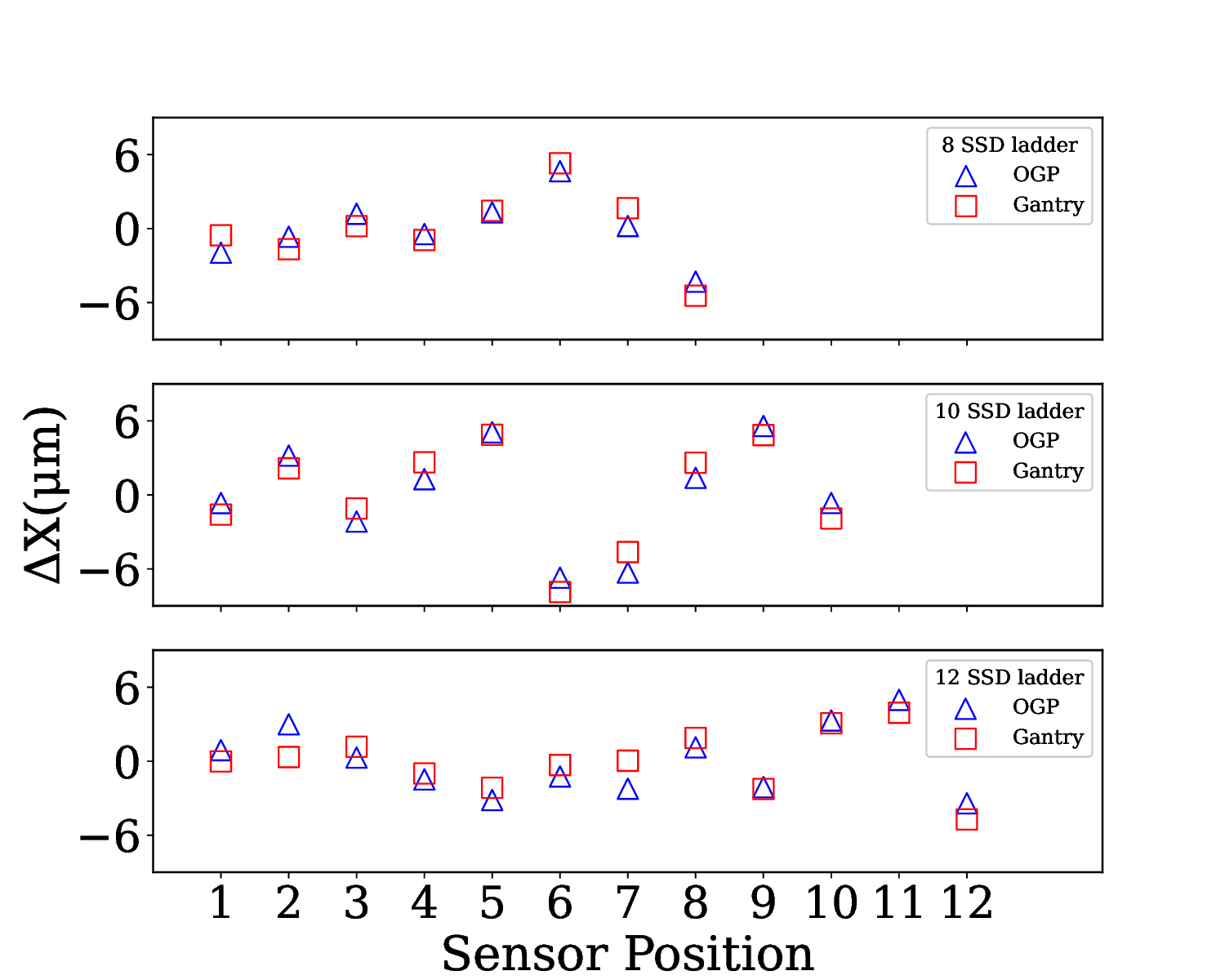}
    \caption{Comparison of gantry and OGP results}
    \label{fig:ganrtry-OGP}
\end{figure}

The alignment accuracy of the assembled ladder was measured using an optical measuring instrument (Mitutoyo QV-X606P3L-D), so-called OGP system and equipped with an optical microscope, offering a three-dimensional precision of (1.6 + 5L/1000) $\mu m$, where L is the measurement length in millimeters. Furthermore, since the gantry system itself incorporates machine vision capabilities, it can also be used to inspect the assembly accuracy of the ladder upon completion. A comparison of measurement results for the same ladder obtained from both the gantry and the OGP system showed a deviation of better than 3 $\mu m$. In Fig. \ref{fig:ganrtry-OGP}, the results from top to bottom correspond to the 8-SSD ladder, 10-SSD ladder, and 12-SSD ladder, with active lengths of 64 $cm$, 80 $cm$, and 96 $cm$, respectively.

\section{Conclusion}
\label{sec5}
\begin{figure}[t]
    \centering
    \begin{subfigure}{0.3\textwidth}
        \includegraphics[width=\linewidth]{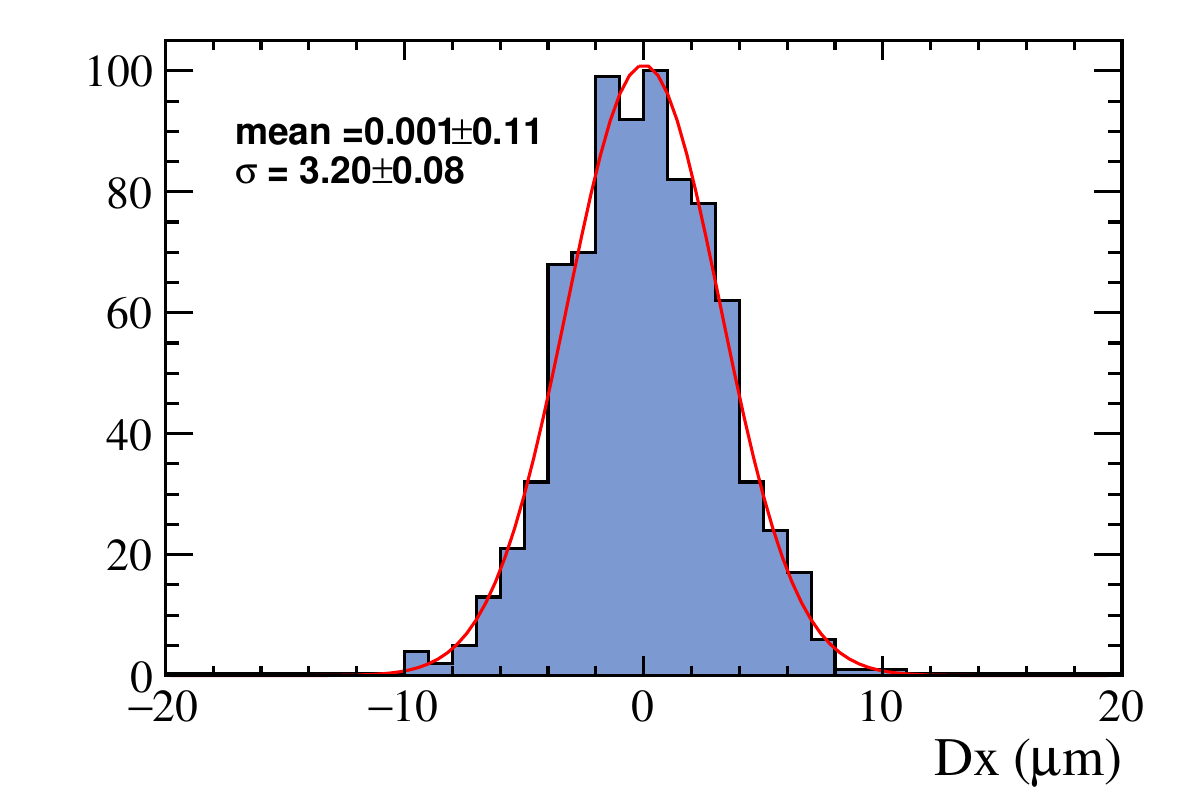}
        \caption{}
        \label{fig:alignment_Dx}
    \end{subfigure}
     \begin{subfigure}{0.3\textwidth}
        \includegraphics[width=\linewidth]{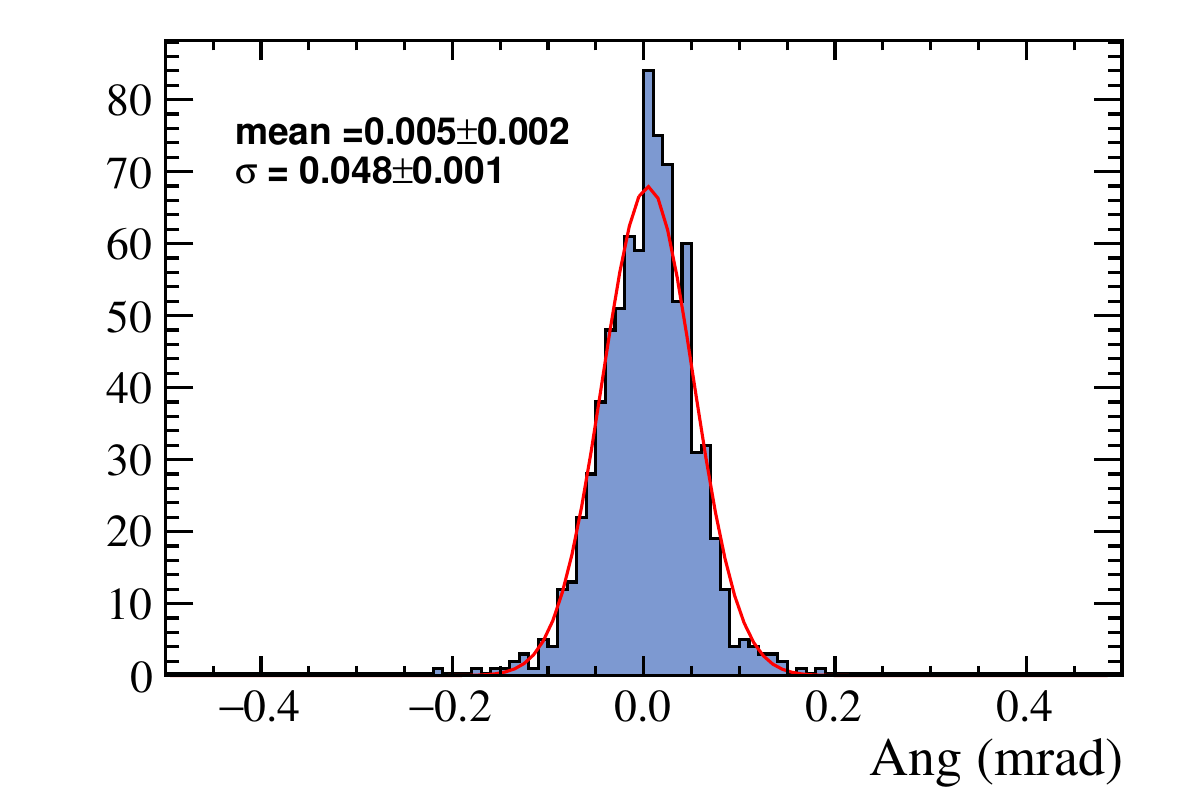}
        \caption{}
        \label{fig:alignment_Angle}
    \end{subfigure}
     \begin{subfigure}{0.3\textwidth}
        \includegraphics[width=\linewidth]{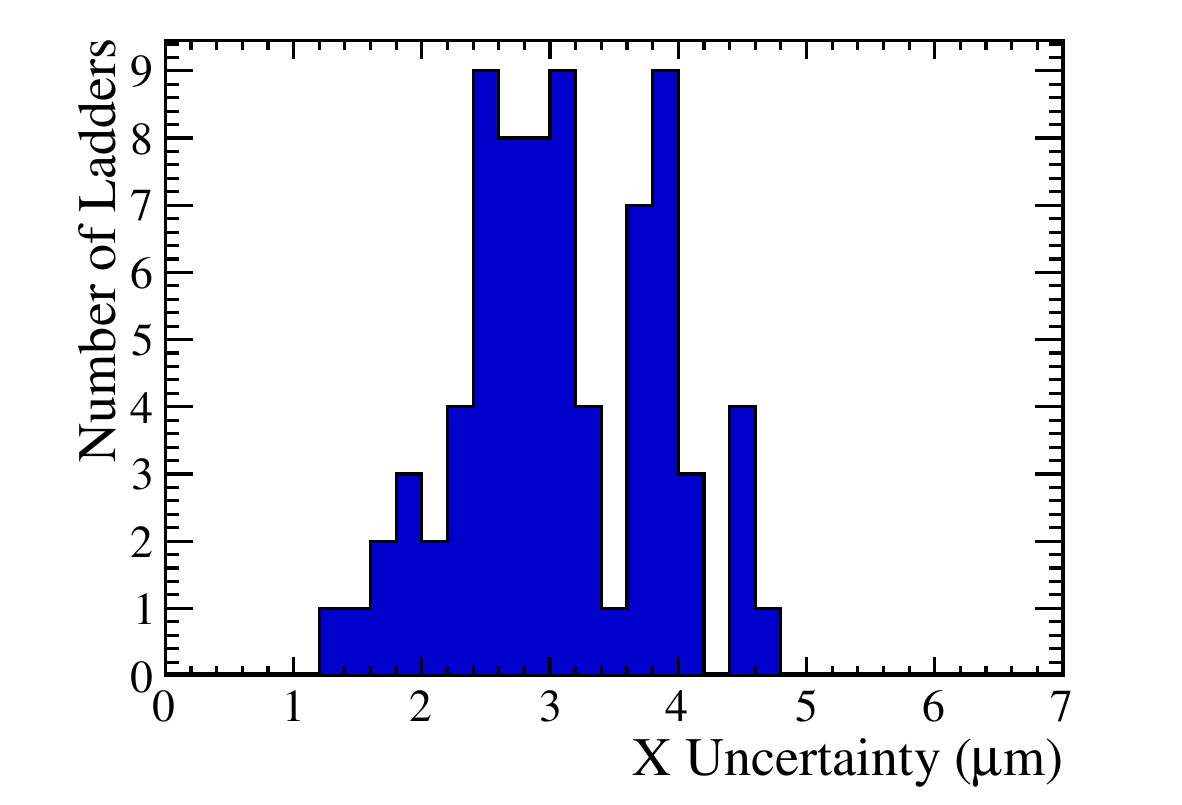}
        \caption{}
        \label{fig:alignment_Ladder}
    \end{subfigure}
    \caption{(a) The distribution of the difference between the individual sensor positions and the center line parallel to the ladder length, so-called $Dx$; (b) The distribution of the U-axis rotation angle for individual sensor relative to the center line of each ladder; (c) The overall alignment accuracy along the strip direction for each individual ladder.}
\end{figure}

Based on the aforementioned process, we have successfully produced a total of 72 flight-model ladders, comprising 16 units of 8-SSD ladders, 16 units of 10-SSD ladders, and 40 units of 12-SSD ladders. The distributions of the X axis position accuracy and the U axis rotation accuracy are shown in Fig \ref{fig:alignment_Dx} and \ref{fig:alignment_Angle}, measured by the OGP system as a crosscheck. The alignment precision is defined as the combination of the difference between the individual sensor positions and the line which defines the axis parallel to the ladder length, and the corrections from the U-axis rotation angle. Therefore, the overall alignment precision is calculated as
\begin{equation}
    \sigma_{\mbox{tot}} = \sqrt{\sigma^2_{Dx}+\frac{(\sigma_\theta\cdot L)^2}{n}}=3.4~\mu m,
\end{equation}
where $n$ represents the number of SSDs for each type of the ladder. The alignment accuracy along strip direction of each individual ladder is shown in Fig \ref{fig:alignment_Ladder}, for the total 72 flight-model ladders and another 4 spear ladders, where all the ladders have the alignment accuracy better than 5 $\mu m$.

A comparison with the assembly precision of other detectors—including the Micro-vertex Detector at LEP (SMD) \cite{Santocchia:1994ce, L3SMD:1994qru}, AMS-01 \cite{Burger:1999sb, Alpat:2000xz}, and AMS-02 \cite{Burger:2007zz} is summarized in Table \ref{tab:Alignment}. It is noteworthy that the SMD, AMS-01 and AMS-02 ladder precision primarily reflects the mechanical precision of the construction jigs and the cutting precision of the wafers. In our approach, the assembly accuracy does not rely on mechanical machining precision due to the adoption of optical positioning. For a 60 $cm$ long ladder, this method achieves an alignment accuracy of up to 2.7 $\mu m$, while ensuring an alignment accuracy better than 4 $\mu m$ for longer ladders (achieving 3.3 $\mu m$ over a length of 1 meter).

\begin{table}
\centering
\begin{tabular}{l|l|l|l|l r}
 & No. of sensor in one ladder & active length cm & Assembly precision ($\mu m$)&$\Delta X$/length\\\hline
 & 8  & 64 & 2.7 & 4.22$\times 10^{-9}$\\
AMS-L0 & 10  & 80 & 3.3 &4.13$\times 10^{-9}$\\
 & 12  & 96 & 3.3 & 3.44$\times 10^{-9}$ \\\hline
AMS-02& 7--15  & 30--60 & 4.6 & 7.67$\times 10^{-9}$\\
AMS-01& 7--15  & 30--60 & 4.7 & 7.83$\times 10^{-9}$\\
SMD & 4  & 28 & 8.2&2.92$\times 10^{-8}$\\
\end{tabular}
\caption{\label{tab:Alignment}Assembly accuracy of AMS-L0 compared to AMS-01, AMS-02 and L3-SMD}
\end{table}

Furthermore, the gantry system utilizes automated control and incorporates real-time monitoring of sensor placement accuracy through machine vision. It has significantly increased the production yield of ladder mass production. This detector construction technology also provides a solution for future large-area with high-precision silicon detector production.

\section*{Acknowledgments}
We acknowledge the enduring support provided by the following funding agencies: the Ministry of Science and Technology of China (MOST) under Grant [2022YFA1604801]. The authors would like to acknowledge the support of the CMS HGCal team at IHEP for utilizing their gantry to complete the construction of the mechanical ladder.

\end{document}